\documentclass[twocolumn,showpacs,preprintnumbers,amsmath,amssymb,nofootinbib]{revtex4-1}
\usepackage{graphicx}
\usepackage{dcolumn}
\usepackage{bm}
\usepackage{mathtools}
\begin{document}

\title{Circuit-model formulas for external-Q factor of resonant cavities \\with capacitive and inductive coupling}
\author{Osamu Kamigaito}
\email{kamigait@riken.jp}
\affiliation{RIKEN Nishina Center for Accelerator-Based Science\\
2-1 Hirosawa, Wako-shi, Saitama 351-0198, Japan}%

\begin{abstract}
The external-Q factor of a resonant mode of a cavity represents the strength of the 
electromagnetic coupling between this resonant mode and the transmission mode of a waveguide coupled to the cavity.
In this paper, we derive formulas to explicitly give the external Q of the cavities coupled with the waveguide
through capacitive and inductive coupling.
The derivation is based on the classical method that uses the reflection coefficient to estimate the coupling strength.
Although the external Q is evaluated via three-dimensional computer simulations currently, 
these formulas may be useful for making speculations in the initial stages of cavity design.

\end{abstract}


\maketitle

\section{Introduction}

The external-Q factor is a fundamental parameter of the cavity resonator; it determines the electromagnetic power 
radiated from the cavity to the waveguide.
In the case of superconducting cavities, the external Q is especially important because
the internal loss of rf power is negligible in comparison to the radiated power, 
and the bandwidth of the rf system is determined by the external Q.
Therefore, it is essential to estimate the external Q while designing the cavity and rf system.

The standard method to evaluate the external Q is described in reference {\cite{wei}}.
This method uses the rf power radiated to the waveguide in addition to the cavity parameters
to evaluate the external Q and coupling coefficient.
It also provides an equivalent circuit of the cavity coupled with the waveguide.
This treatment is widely used for the analysis of the cavity voltage because
it precisely describes the electromagnetic field in the cavity. 

However, the aforementioned method has the following two problems.
First, this method only yields an equivalent circuit coupled with an ideal transformer.
Therefore, the properties of the actual circuit, such as the input impedance and reflection coefficient, 
are not correctly represented.
Second, the external Q can be obtained if only the input impedance, observed from the waveguide, is known{\cite{kam}}.
In other words, the radiation power is not required to obtain the external Q.

A general method to calculate the external Q using the input impedance is given in reference {\cite{kam}}, which,
however, requires numerical calculation.
Therefore, in this paper, we derive an explicit formulas to determine the external Q by another approach based on the reflection coefficient.
We will see that the external Q are expressed in terms of the cavity parameters,
through calculations of the refection coefficient for the two commonly used coupling schemes, 
i.e., capacitive and inductive coupling.

\section{Reflection coefficient and external Q}

\begin{figure}[htbp]
   \centering
   \includegraphics[width=2.7in]{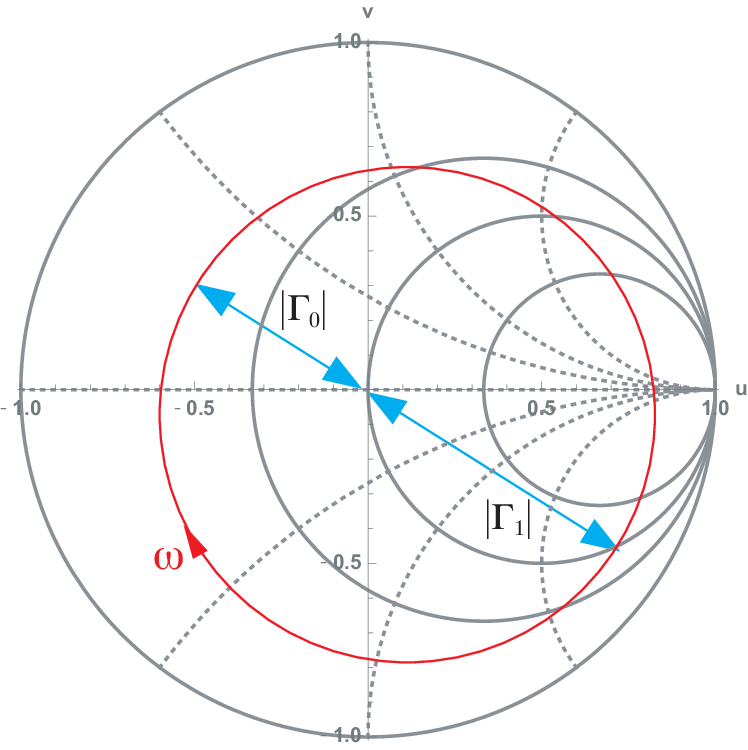} 
   \caption{Reflection coefficient $\Gamma$ of a resonant cavity coupled with a waveguide 
   represented on a Smith chart.
   The minimum and maximum absolute values of the reflection coefficient are denoted by
   $\left|\Gamma_0 \right|$ and $\left|\Gamma_1 \right|$, respectively.}
   \label{fig:gammageneral}
\end{figure}

Let us suppose a general condition where a resonant cavity is coupled with a waveguide.
The reflection coefficient $\Gamma$ represents a circle in the Smith chart around the resonance frequency, 
as shown in Fig. {\ref{fig:gammageneral}.
We use this circle to obtain the coupling coefficient $\beta$, which is defined 
by the ratio of the unloaded Q to the external Q, i.e.,
\begin{eqnarray}
\label{eq:betadef}
\beta\coloneqq\frac{Q_0}{Q_{\mbox {\tiny ext}}}.
\end{eqnarray}

To this end, let the minimum and maximum absolute values of the reflection coefficient be $\left|\Gamma_0 \right|$
and $\left|\Gamma_1 \right|$, respectively, as shown in Fig. {\ref{fig:gammageneral}.
Note that the points corresponding to the minimum and maximum reflection coefficients are at opposite positions
across the origin of the complex plane because $\Gamma$ is a circle.
Then, the voltage standing wave ratios corresponding to the minimum and maximum values of $\left| \Gamma \right|$ are 
given by
\begin{eqnarray}
\label{eq:sigma0}
\sigma_0=\frac{1+\left|\Gamma_0 \right|}{1-\left|\Gamma_0 \right|}
\end{eqnarray}
and
\begin{eqnarray}
\label{eq:sigma1}
\sigma_1=\frac{1+\left|\Gamma_1 \right|}{1-\left|\Gamma_1 \right|},
\end{eqnarray}
respectively.
According to reference \cite{sla}, the 
coupling coefficient $\beta$ can be given by
\begin{eqnarray}
\label{eq:beta}
\beta = \left\{ \begin{array}{l}
\displaystyle\sigma_0-1/\sigma_1 \quad ({\mbox {in the case of overcoupling}}), \\ \\
1/\sigma_0-1/\sigma_1 \quad ({\mbox {in the case of undercoupling}}).
\end{array} \right.
\end{eqnarray}
Therefore, it is possible to obtain $Q_{\mbox {\tiny ext}}$ once $\left|\Gamma_0 \right|$ 
and $\left|\Gamma_1 \right|$ are known
for a given cavity coupled with a waveguide.

\section{Capacitive coupling}

\begin{figure}[htbp]
   \centering
   \includegraphics[width=2.7in]{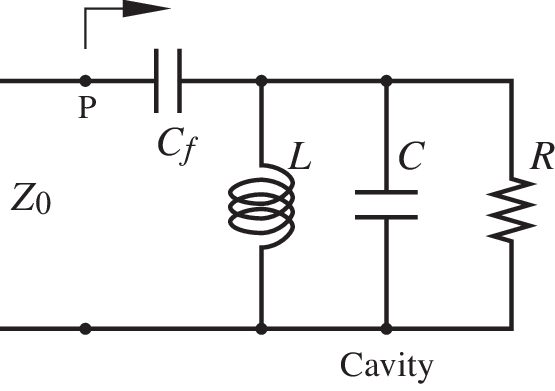} 
   \caption{Circuit diagram representing a capacitively coupled cavity.}
   \label{fig:c}
\end{figure}

Consider an LCR parallel resonant circuit coupled with a coaxial waveguide with 
a characteristic 
impedance $Z_0$ through the coupling capacitance $C_f$, as shown in Fig. \ref{fig:c}.
It represents an equivalent circuit of a cavity coupled capacitively with the waveguide.
The resonance frequency and 
unloaded Q of the cavity are related to the circuit parameters as follows, respectively:
\begin{eqnarray}
\omega_r&=&\frac{1}{\sqrt{L C}},\\
\label{eq:defq0}
Q_0&=&\omega_r  C  R.
\end{eqnarray}
In this paper, we assume that
\begin{eqnarray}
\label{eq:zentei}
Z_0 \ll R.
\end{eqnarray}

The input impedance of the cavity at point P in Fig. \ref{fig:c} is given by the following equation.
\begin{eqnarray}
Z_{\mbox {\tiny in}}\left(\omega\right)&=&\frac{1}{j\omega C_f}+\frac{1}{1/R+j\omega C+1/(j\omega L)}\nonumber \\
&=&\frac{1}{j\omega C_f}+\frac{R}{1+j Q_0 \delta(\omega)}.
\label{eq:zin}
\end{eqnarray}
Here, we set
\begin{eqnarray}
\delta(\omega)\coloneqq\frac{\omega}{\omega_r }-\frac{\omega_r }{\omega}.
\label{eq:delta}
\end{eqnarray}
Let us define the second term of the last line of Eq. (\ref{eq:zin}) by
\begin{eqnarray}
Z_p\coloneqq\frac{R}{1+j Q_0 \delta(\omega)}.
\label{eq:zp}
\end{eqnarray}
This denotes the impedance of the parallel resonant circuit.

In this paper, the resonant cavities are treated with a very large $Q_0$, such as the superconducting cavities.
In this case, it is sufficient to only consider the neighborhood of the resonance frequency;
thus, the slowly changing factor $\omega$ 
in the first term of the right-hand side of Eq. (\ref{eq:zin})
may be replaced with $\omega_r$.
Therefore, we consider the following equation instead of Eq. (\ref{eq:zin}).
\begin{eqnarray}
Z_a=\frac{1}{j\omega_r C_f}+\frac{R}{1+j \Delta}.
\label{eq:za}
\end{eqnarray}
Here, we set
\begin{eqnarray}
\Delta\coloneqq Q_0 \delta(\omega).
\label{eq:Delta}
\end{eqnarray}

The real and imaginary parts of Eq. (\ref{eq:za}) are written as follows, respectively:
\begin{eqnarray}
\label{eq:reza}
\Re{Z_a}&=&\frac{R}{1+\Delta^2}\\
&=&\Re{Z_p},\\
\Im{Z_a}&=&-\frac{R \Delta}{1+\Delta^2}-\frac{1}{\omega_r C_f}\\
\label{eq:imza}
&=&\Im{Z_p}-\frac{1}{\omega_r C_f}.
\end{eqnarray}
The implications of these expressions are shown in Fig. \ref{fig:reimza},
which plots $Z_a$ and $Z_p$ against $\Delta$.
It is evident that the real part of $Z_a$ coincides with the real part of $Z_p$.
In contrast, the imaginary part of $Z_a$ is shifted downward 
from the imaginary part of $Z_p$ by $1/\omega_r C_f$.

\begin{figure}[htbp]
   \centering
   \includegraphics[width=3.4in]{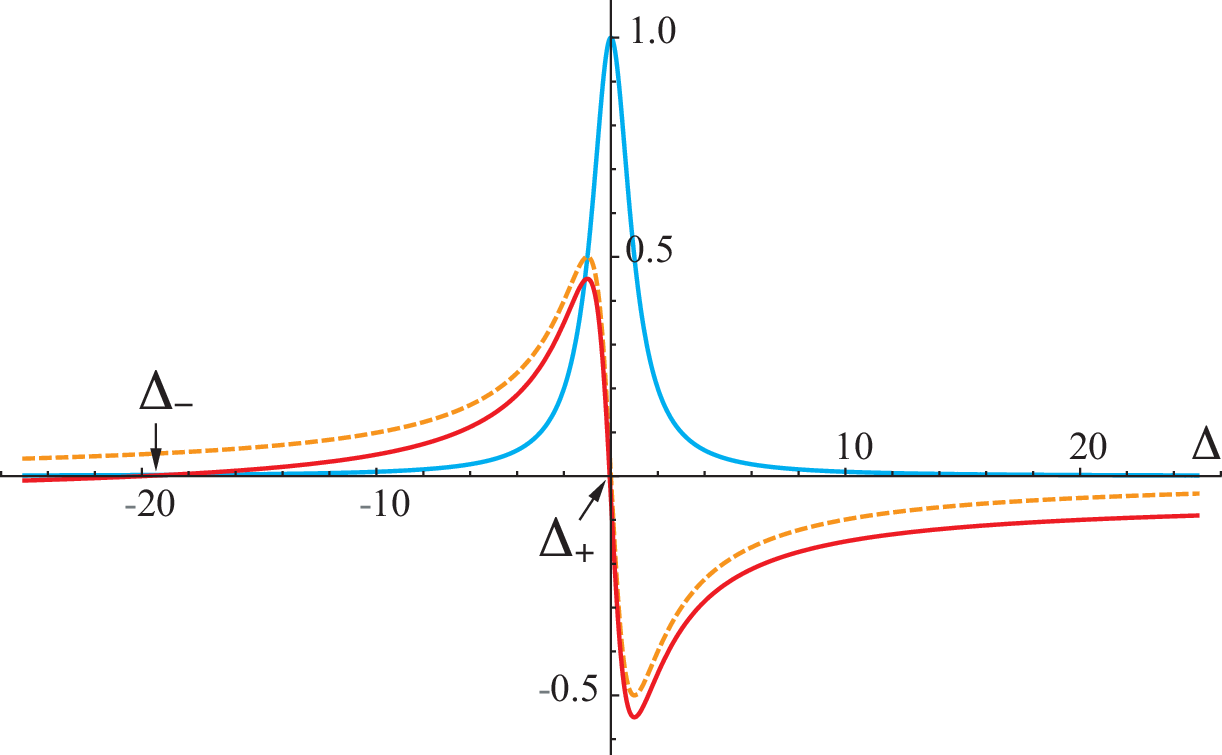} 
   \caption{Conceptual diagram of $Z_a$ and $Z_p$ plotted as functions of $\Delta$.
   The blue curve represents the real part of $Z_a$, which is identical to the real part of $Z_p$.
   The red curve represents the imaginary part of $Z_a$, which is obtained by
   shifting the orange dashed curve downward, which represents the imaginary part of $Z_p$, 
   by $1/\omega_r C_f$.
   Note that the equation $\Im{Z_a}=0$ always has two roots, denoted
   by $\Delta_-$ and $\Delta_+$, where $\Delta_- < \Delta_+$.}
   \label{fig:reimza}
\end{figure}

Here, we {\bf assume} that $1/\omega_r C_f$ is very small in comparison to $R$, 
i.e.\footnote{We will see in the Appendix that this condition is fulfilled when Eq. (\ref{eq:zentei}) is valid.},
\begin{eqnarray}
\omega_r C_f R \gg 1.
\label{eq:zure}
\end{eqnarray}
Under this condition, the equation $\Im{Z_a}=0$ or its equivalent quadratic equation,
\begin{eqnarray}
\Delta^2+\omega_r C_f R \Delta +1 =0,
\label{eq:quadc}
\end{eqnarray}
has two roots, $\Delta_-$ and $\Delta_+$, as schematically shown in Fig. \ref{fig:reimza}, 
where they are defined as $\Delta_- < \Delta_+$.
Therefore, the following relationships are valid.
\begin{eqnarray}
\label{eq:delwa}
\Delta_- +\Delta_+ &=& -\omega_r C_f R,\\
\label{eq:delseki}
\Delta_-\cdot\Delta_+ &=&1.
\end{eqnarray}
It should be noted here that 
the input impedance $Z_a$ is real at $\Delta = \Delta_-$ and $\Delta = \Delta_+$.
These real values at $\Delta_-$ and $\Delta_+$ are denoted by $Z_-$ and $Z_+$, respectively.
It is evident that the following relationships are also valid considering
Eqs. (\ref{eq:reza}) and (\ref{eq:quadc}).
\begin{eqnarray}
Z_- + Z_+ &=&R,\\
Z_-\cdot Z_+ &=&\frac{1}{\left(\omega_r C_f\right)^2}.
\end{eqnarray}

The value of $\Delta_+$ is approximately zero, as shown in Fig. \ref{fig:reimza}.
Therefore, from Eqs.  (\ref{eq:delwa}) and (\ref{eq:delseki}), we have 
\begin{eqnarray}
\Delta_- &\approx& -\omega_r C_f R
\end{eqnarray}
and
\begin{eqnarray}
\Delta_+&\approx&-\frac{1}{\omega_r C_f R}\quad(\approx 0);
\end{eqnarray}
correspondingly\footnote{We will consider the impedance matching condition in the Appendix.}, 
\begin{eqnarray}
\label{eq:zminus}
Z_- &\approx& \frac{1}{\left(\omega_r C_f\right)^2 R}
\end{eqnarray}
and
\begin{eqnarray}
\label{eq:zplus}
Z_+&\approx&R.
\end{eqnarray}

It is evident from Eq. (\ref{eq:zminus}) and assumption (\ref{eq:zure}) that 
\begin{eqnarray}
\omega_r C_f Z_- \approx \frac{1}{\omega_r C_f R} \ll 1.
\end{eqnarray}
Moreover, because the orders of $Z_-$ and $Z_0$ are similar to each other in the usual cavities, 
the following expression is also valid with a good approximation.
\begin{eqnarray}
\label{eq:kinji}
\omega_r C_f Z_0 \ll 1.
\end{eqnarray}

\begin{figure}[htbp]
   \centering
   \includegraphics[width=3in]{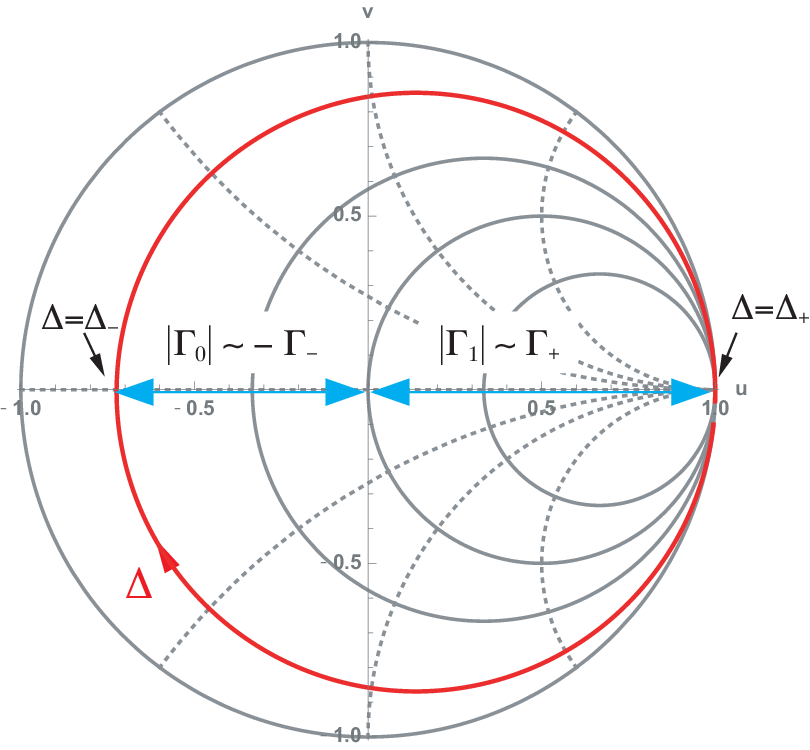} 
   \caption{Reflection coefficient of the capacitively coupled resonant circuit under the overcoupling condition.}
   \label{fig:gammaza}
\end{figure}

Now, let us consider the reflection coefficient $\Gamma$ of this circuit as observed 
from point P in Fig. \ref{fig:c}
using the conditions obtained above.
First, note that $Y_p\coloneqq 1/Z_p$ is a straight line in the complex $Y$-plane.
This straight line $Y_p$ is mapped to $\Gamma$ with two successive linear fractional transformations,
namely, $Z_a=1/Y_p+1/j\omega_r C_f$ and $\Gamma=(Z_a-Z_0)/(Z_a+Z_0)$.
Therefore, $\Gamma$ certainly becomes a circle in the complex $\Gamma$-plane.
Second, by setting the reflection coefficient at $\Delta=\Delta_+$ as $\Gamma_+$, we obtain 
the following approximation using Eqs. (\ref{eq:zplus}) and (\ref{eq:zentei}).
\begin{eqnarray}
\Gamma_+&=&\frac{Z_+-Z_0}{Z_++Z_0}\nonumber\\
&\approx&\frac{R-Z_0}{R+Z_0}\nonumber\\
&\approx& 1.
\end{eqnarray}
In contrast, Fig. \ref{fig:reimza} shows that $Z_a$ becomes approximately $1/j\omega_r C_f$ 
when $\omega$ moves away from $\omega_r$ to some extent.
Thus, using Eq. (\ref{eq:kinji}), the approximate value of the 
corresponding refection coefficient $\Gamma_{\infty}$ 
can be given by
\begin{eqnarray}
\Gamma_{\infty}&=&\frac{\frac{1}{j\omega_r C_f}-Z_0}{\frac{1}{j\omega_r C_f}+Z_0}\nonumber\\
&\approx& 1.
\end{eqnarray}
In other words,  
the point with the largest reflection coefficient in this circuit 
is very close to the point corresponding to $\Delta=\Delta_+$,
which lies on the real axis, as shown in Fig. \ref{fig:gammaza}.
Therefore, the following value may be used:
\begin{eqnarray}
\left|\Gamma_1 \right|=1.
\label{eq:gamma1}
\end{eqnarray}
Consequently, the point with the smallest reflection coefficient should be very close to the
point corresponding to $\Delta=\Delta_-$,
which is on the real axis, as shown in Fig. \ref{fig:gammaza}.
In other words, the following approximation may be used:
\begin{eqnarray}
\Gamma_0\approx\Gamma_-=\frac{Z_--Z_0}{Z_-+Z_0}.
\end{eqnarray}
This implies that
\begin{eqnarray}
\left|\Gamma_0 \right| = \left\{ \begin{array}{l}
\displaystyle\frac{Z_0-Z_-}{Z_-+Z_0} \quad ({\mbox {in the case of overcoupling}}), \\ \\
\displaystyle\frac{Z_--Z_0}{Z_-+Z_0} \quad ({\mbox {in the case of undercoupling}}).
\end{array} \right.
\end{eqnarray}

Thus, owing to the aforementioned results, the following approximated expressions 
can be obtained for the voltage standing wave ratio,
\begin{eqnarray}
1/\sigma_1&=&0,\\
\sigma_0 &=& \left\{ \begin{array}{l}
\displaystyle\frac{Z_0}{Z_-} \quad ({\mbox {in the case of overcoupling}}), \\ \\
\displaystyle\frac{Z_-}{Z_0} \quad ({\mbox {in the case of undercoupling}}).
\end{array} \right.
\end{eqnarray}
Finally, we obtain the coupling coefficient by substituting these values in Eq. (\ref{eq:beta})
as follows:
\begin{eqnarray}
\beta=\frac{Z_0}{Z_-}=Z_0(\omega_r C_f)^2 R.
\end{eqnarray}
This expression is valid irrespective of the coupling strength.
Moreover, owing to Eq. (\ref{eq:betadef}), the external Q can be given by
\begin{eqnarray}
Q_{\mbox {\tiny ext}}=\frac{C}{Z_0 \omega_r {C_f}^2}.
\end{eqnarray}
It should be noted is that the coupler tip (point P in Fig. \ref{fig:c}) corresponds to the short end 
at the operating frequency when $\beta\gg 1$, as shown in Fig. \ref{fig:gammaza}.

\section{Inductive coupling}

\begin{figure}[htbp]
   \centering
   \includegraphics[width=2.7in]{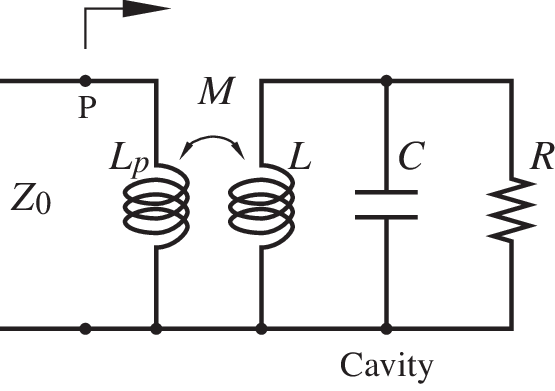} 
   \caption{Circuit diagram representing an inductively coupled cavity.}
   \label{fig:loop}
\end{figure}

The cavities with inductive coupling can be treated in almost the same manner as 
those with capacitive coupling. 
Let us consider an inductively coupled circuit, as shown in Fig. \ref{fig:loop},
considering Eq. (\ref{eq:zentei}).
The input impedance obsereved from point P in Fig. \ref{fig:loop} leads to
the following expression:
\begin{eqnarray}
\label{eq:zalorig}
Z_{\mbox {\tiny in}}\left(\omega\right)
=j\omega L_p + \frac{R}{\omega_r^2 L^2}\cdot \frac{\omega^2 M^2}{1+j Q_0 \delta},
\end{eqnarray}
where $Q_0$ and $\delta$ are given by Eqs. (\ref{eq:defq0}) and (\ref{eq:delta}), respectively.

Because our consideration is limited in the neighborhood of $\omega_r$, 
similar to the case of capacitively coupled circuit, 
we consider the following expression,
where $\omega$ is replaced with $\omega_r$ in the first term of the right hand side of Eq. (\ref{eq:zalorig}):
\begin{eqnarray}
\label{eq:zal}
Z_a\left(\omega\right)
=j\omega_r L_p + \frac{M^2}{L^2}\cdot \frac{R}{1+j \Delta},
\end{eqnarray}
Note that the term $\Delta$ is given by Eq. (\ref{eq:Delta}).
The real and imaginary parts of $Z_a$ in Eq. (\ref{eq:zal}) are respectively given as follows:
\begin{eqnarray}
\label{eq:rezal}
\Re{Z_a}&=& \frac{M^2}{L^2}\cdot\frac{R}{1+\Delta^2},\\
\label{eq:imzal}
\Im{Z_a}&=&-\frac{M^2}{L^2}\cdot\frac{R \Delta}{1+\Delta^2}+\omega_r L_p.
\end{eqnarray}
Figure \ref{fig:reimzal} illustrates the implications of Eqs. (\ref{eq:rezal}) and (\ref{eq:imzal}).
\begin{figure}[htbp]
   \centering
   \includegraphics[width=3.4in]{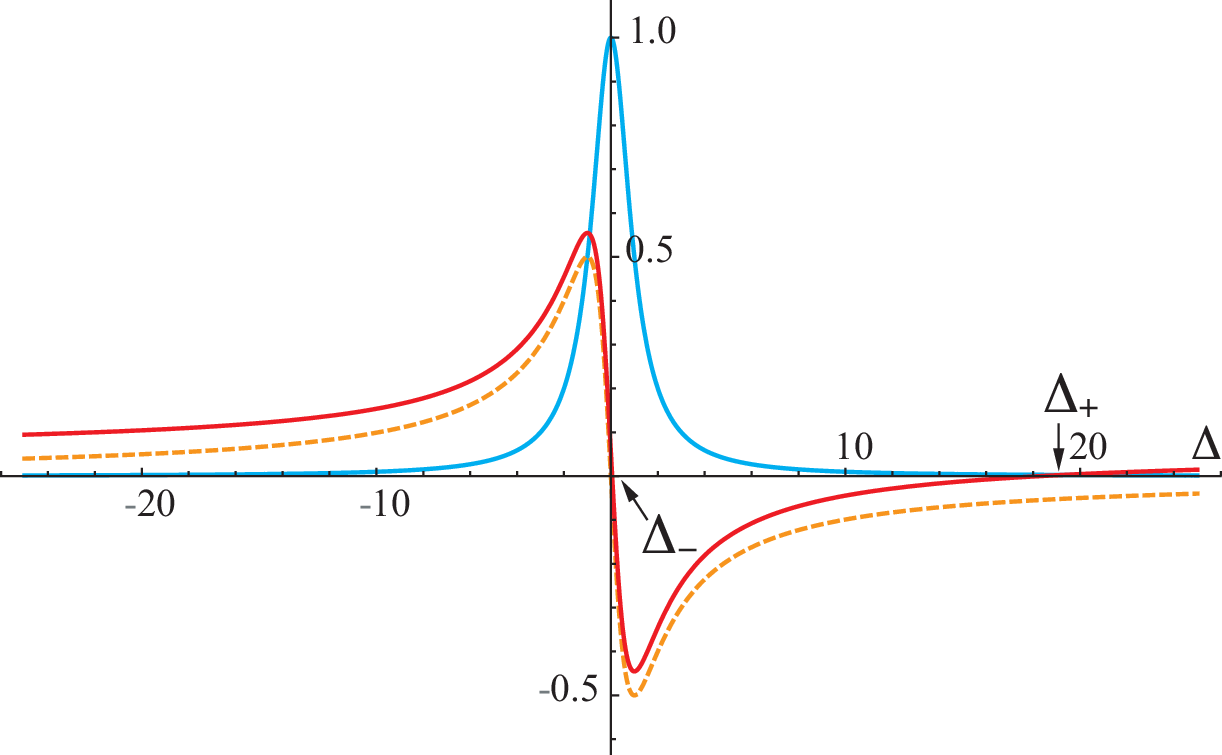} 
   \caption{Conceptual diagram of $Z_a$ and $Z_p$ plotted as functions of $\Delta$.
   The blue curve represents the real part of $Z_a$, which is identical to the real part of $Z_p$.
   The red curve represents the imaginary part of $Z_a$, which is obtained by
   shifting the orange dashed curve upward, which represents the imaginary part of $Z_p$,
   by $\omega_r L_p$.
   Note that the equation $\Im{Z_a}=0$ always has two roots, denoted
   by $\Delta_-$ and $\Delta_+$, where $\Delta_- < \Delta_+$.}
   \label{fig:reimzal}
\end{figure}

Here, we {\bf assume} that $\omega_r L_p$ is very small in comparison to $(M/L)^2R$, i.e.,
\begin{eqnarray}
\omega_r L_p \ll (M/L)^2R.
\label{eq:zurel}
\end{eqnarray}
Using the following expression,
\begin{eqnarray}
k^2\coloneqq\frac{M^2}{L L_p},
\end{eqnarray}
this condition can also be written as\footnote{This condition will be examined in the Appendix.}
\begin{eqnarray}
k^2 Q_0 \gg 1.
\label{eq:zurel2}
\end{eqnarray}
Under this condition, the equation $\Im{Z_a}=0$ has two roots, as shown in Fig. \ref{fig:reimzal}.
We set the roots as $\Delta_-$ and $\Delta_+$, where $\Delta_- < \Delta_+$.
They denote the solutions of the following quadratic equation:
\begin{eqnarray}
\Delta^2-k^2 Q_0 \Delta +1 =0.
\label{eq:quadl}
\end{eqnarray}
Therefore, the following relationships are valid.
\begin{eqnarray}
\label{eq:delwal}
\Delta_- +\Delta_+ &=& k^2 Q_0
\end{eqnarray}
and
\begin{eqnarray}
\label{eq:delsekil}
\Delta_-\cdot\Delta_+ &=&1.
\end{eqnarray}
In contrast, the input impedance $Z_a$ is real at $\Delta = \Delta_-$ and $\Delta = \Delta_+$.
We denote these real values at $\Delta_-$ and $\Delta_+$ by $Z_-$ and $Z_+$, respectively.
It is evident that the following relationships are also valid when we consider 
Eqs. (\ref{eq:rezal}) and (\ref{eq:quadl}):
\begin{eqnarray}
Z_- + Z_+ &=&\frac{M^2}{L^2}R,\\
Z_-\cdot Z_+ &=&\left(\omega_r L_p\right)^2.
\end{eqnarray}

The value of $\Delta_-$ is approximately zero, as shown in Fig. \ref{fig:reimzal}.
Therefore, from Eqs.  (\ref{eq:delwal}) and (\ref{eq:delsekil}),
\begin{eqnarray}
\Delta_+ &\approx& k^2 Q_0
\end{eqnarray}
and
\begin{eqnarray}
\Delta_-&\approx&\frac{1}{k^2 Q_0}\quad(\approx 0);
\end{eqnarray}
correspondingly\footnote{We will consider the impedance matching condition in the Appendix.},
\begin{eqnarray}
\label{eq:zplusl}
Z_+ &\approx& \frac{\omega_r L_p}{k^2 Q_0}
\end{eqnarray}
and
\begin{eqnarray}
\label{eq:zminusl}
Z_-&\approx&\frac{M^2}{L^2}R.
\end{eqnarray}

It can be observed from Eq. (\ref{eq:zminusl}) and assumption (\ref{eq:zurel2}) that 
\begin{eqnarray}
\frac{Z_+}{\omega_r L_p} \approx \frac{1}{k^2 Q_0} \ll 1.
\end{eqnarray}
Moreover, because the orders of $Z_+$ and $Z_0$ are similar to each other in the usual cavities, 
the following expression also holds true with a good approximation:
\begin{eqnarray}
\frac{Z_0}{\omega_r L_p} \ll 1.
\label{eq:kinjil}
\end{eqnarray}

\begin{figure}[htbp]
   \centering
   \includegraphics[width=3in]{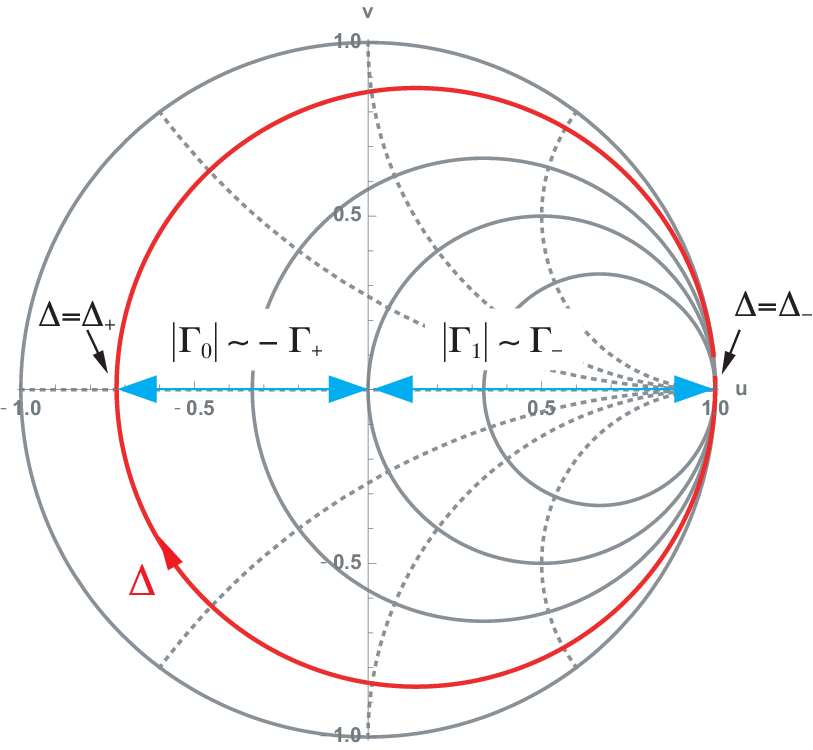} 
   \caption{Reflection coefficient of the inductively coupled resonant circuit under the overcoupling condition.}
   \label{fig:gammazal}
\end{figure}

By setting the reflection coefficient at $\Delta=\Delta_-$ as $\Gamma_-$, 
we obtain the following approximation using Eqs. (\ref{eq:zplusl}), (\ref{eq:zurel}), and (\ref{eq:kinjil}):
\begin{eqnarray}
\Gamma_-&=&\frac{Z_--Z_0}{Z_-+Z_0}\nonumber\\
&\approx&\frac{(M/L)^2R-Z_0}{(M/L)^2R+Z_0}\nonumber\\
&\approx& 1.
\end{eqnarray}
In contrast, Fig. \ref{fig:reimzal} shows that $Z_a\approx j\omega_r L_p$ 
when $\omega$ moves away from $\omega_r$ to some extent.
Thus, using Eq. (\ref{eq:kinjil}), the approximate value of the corresponding 
refection coefficient $\Gamma_{\infty}$ can be given by
\begin{eqnarray}
\Gamma_{\infty}&=&\frac{ j\omega_r L_p-Z_0}{ j\omega_r L_p+Z_0}\nonumber\\
&\approx& 1.
\end{eqnarray}
In other words,  
the point with the largest reflection coefficient in this circuit 
is very close to the point corresponding to $\Delta=\Delta_-$,
which lies on the real axis, as shown in Fig. \ref{fig:gammazal}.
Therefore, the following value may be used:
\begin{eqnarray}
\left|\Gamma_1 \right|=1.
\label{eq:gamma1}
\end{eqnarray}
Consequently, the point with the smallest reflection coefficient should be very close to the
point corresponding to $\Delta=\Delta_+$,
which is on the real axis, as shown in Fig. \ref{fig:gammazal}.
In other words, the following approximation may be used:
\begin{eqnarray}
\Gamma_0\approx\Gamma_+=\frac{Z_+-Z_0}{Z_++Z_0}.
\end{eqnarray}
This implies that
\begin{eqnarray}
\left|\Gamma_0 \right| = \left\{ \begin{array}{l}
\displaystyle\frac{Z_0-Z_+}{Z_++Z_0} \quad ({\mbox {in the case of overcoupling}}), \\ \\
\displaystyle\frac{Z_+-Z_0}{Z_++Z_0} \quad ({\mbox {in the case of undercoupling}}).
\end{array} \right.
\end{eqnarray}

Owing to these results, the following approximations can be obtained 
for the voltage standing wave ratio:
\begin{eqnarray}
1/\sigma_1&=&0,\\
\sigma_0 &=& \left\{ \begin{array}{l}
\displaystyle\frac{Z_0}{Z_+} \quad ({\mbox {in case of overcoupling}}), \\ \\
\displaystyle\frac{Z_+}{Z_0} \quad ({\mbox {in case of undercoupling}}).
\end{array} \right.
\end{eqnarray}
Finally, we obtain the coupling coefficient by substituting these values in Eq. (\ref{eq:beta})
as follows:
\begin{eqnarray}
\beta=\frac{Z_0}{Z_+}=\frac{Z_0 k^2 Q_0}{\omega_r L_p}.
\end{eqnarray}
This expression holds true irrespective of the coupling strength.
Moreover, owing to Eq. (\ref{eq:betadef}), the external Q can be given by
\begin{eqnarray}
Q_{\mbox {\tiny ext}}=\frac{\omega_r L_p}{Z_0 k^2}.
\end{eqnarray}
It should be noted that the coupler tip (point P in Fig. \ref{fig:loop}) also corresponds to 
the short end 
at the operating frequency when $\beta\gg 1$, as shown in Fig. \ref{fig:gammazal}.

\section{Summary}

The external Q and coupling coefficient of the capacitively coupled resonant circuit shown in Fig. \ref{fig:c} 
can be represented by
\begin{eqnarray}
\label{eq:kekkaqe}
Q_{\mbox {\tiny ext}}&=&\frac{C}{Z_0 \omega_r {C_f}^2}
\end{eqnarray}
and
\begin{eqnarray}
\label{eq:kekkabeta}
\beta&=&Z_0 (\omega_r C_f)^2 R,
\end{eqnarray}
respectively, 
where $\omega_r$ is the angular resonance frequency of the cavity.

The external Q and coupling coefficient of the inductively coupled resonant circuit shown in Fig. \ref{fig:loop} 
can be represented by
\begin{eqnarray}
\label{eq:kekkaqel}
Q_{\mbox {\tiny ext}}&=&\frac{\omega_r L_p}{Z_0 k^2}
\end{eqnarray}
and
\begin{eqnarray}
\label{eq:kekkabetal}
\beta&=&\frac{Z_0 k^2 Q_0}{\omega_r L_p},
\end{eqnarray}
respectively, 
where $\omega_r$ is the angular resonance frequency of the cavity and $k^2=M^2/L L_p$.

\section{Acknowledgment}

The way to understand the impedance behavior of a resonant circuit on the complex $\Gamma$-plane,
and the graphical representation of the impedance matching were suggested by Dr. Yoshiaki Chiba 
of RIKEN Accelerator Research Facility.
The author would like to thank him for his guidance.

\appendix*
\section{Impedance matching}
\subsection{Capacitive coupling}
First, we determine the impedance matching condition between a cavity and waveguide 
with the characteristic impedance $Z_0$ through capacitive coupling.
It is evident from Eqs. (\ref{eq:reza}) and (\ref{eq:imza}) that 
the matching condition can be obtained by changing $C_f$ such that the following relations may be fulfilled:
\begin{eqnarray}
\label{eq:remat}
\frac{R}{1+\Delta^2}&=&Z_0,\\
\label{eq:immat}
\frac{R \Delta}{1+\Delta^2}+\frac{1}{\omega_r C_f}&=&0.
\end{eqnarray}
By eliminating $\Delta$ from the above equations, we obtain
\begin{eqnarray}
\label{eq:mat}
1+\left(\frac{1}{\omega_r C_f Z_0}\right)^2=\frac{R}{Z_0}.
\end{eqnarray}
This equation can determine the value of $C_f$ in the impedance matching problem.

Equation (\ref{eq:mat}) elucidates that the following 
condition should be maintained for the matching condition when
$R\gg Z_0$:
\begin{eqnarray*}
\omega_r C_f Z_0 \ll 1.
\end{eqnarray*}
This is Eq. (\ref{eq:kinji}).
Under this condition, the first term in the left-hand side of Eq. (\ref{eq:mat}) can be neglected
to obtain
\begin{eqnarray}
\omega_r C_f R \approx \frac{1}{\omega_r C_f Z_0}\gg 1.
\end{eqnarray}
Therefore, Eq. (\ref{eq:zure}) is valid.

\subsection{Inductive coupling}
In the case of inductive coupling, 
the matching condition contains two adjustment parameters, $L_p$ and $M$,
which relatively complicate the problem.
Using Eqs. (\ref{eq:rezal}) and (\ref{eq:imzal}), the 
matching condition can be given by
\begin{eqnarray}
\label{eq:rematl}
\frac{M^2}{L^2}\cdot\frac{R}{1+\Delta^2}&=&Z_0,\\
\label{eq:immatl}
\frac{M^2}{L^2}\cdot\frac{R \Delta}{1+\Delta^2}-\omega_r L_p&=&0.
\end{eqnarray}
By eliminating $\Delta$ from the above equations, we obtain
\begin{eqnarray}
\label{eq:matl}
\frac{L^2}{M^2}\left\{1+\left(\frac{\omega_r L_p}{Z_0}\right)^2\right\}=\frac{R}{Z_0}.
\end{eqnarray}
This equation can determine the value of $L_p$ and $M$ in the impedance matching problem.
To achieve a matching condition when $R\gg Z_0$, 
the following expression must be valid, {\bf unless} $(L/M)^2$ is too large:
\begin{eqnarray*}
\frac{\omega_r L_p}{Z_0} \gg 1.
\end{eqnarray*}
This is Eq. (\ref{eq:kinjil}).
Under these conditions, the first term in the left-hand side of Eq. (\ref{eq:matl}) can be neglected
to obtain
\begin{eqnarray}
k^2 Q_0 \approx \frac{\omega_r L_p}{Z_0} \gg 1.
\end{eqnarray}
This implies that Eq. (\ref{eq:zurel2}) is valid.

The additional condition on $(L/M)^2$ slightly complicates the estimation.
It should be noted that when the value of $(L/M)^2$ is large, 
the value of $\omega_r L_p/Z_0$ given by Eq. (\ref{eq:matl}) is not
correspondingly large.
Under such a condition, Eqs. (\ref{eq:kekkaqel}) and (\ref{eq:kekkabetal}) do not yield good approximations.


\begin{thebibliography}{2}
\bibitem{wei} W. Weingarten,``Superconducting Cavities'', CAS - CERN Accelerator School: RF engineering for particle accelerators, CERN 92-03, p. 318 (1992).
\bibitem{kam} O. Kamigaito,``Circuit-model representation of external-Q calculation'', Phys. Rev. ST Accel. Beams {\bf 9}, 062003 (2006).
\bibitem{sla} J. C. Slater, ``Microwave Electronics'', Rev. Mod. Phys. {\bf 18}, 441 (1946).
Note that the definition of the voltage standing wave ratio in this paper is the inverse of Slater's definition.

\end{thebibliography}
\end{document}